\documentclass[submission,copyright,creativecommons]{eptcs}
\providecommand{\event}{TARK 2015} 

\usepackage{verbatim}
\usepackage[usenames,dvipsnames,svgnames,table]{xcolor}
\usepackage{graphicx}
\usepackage{url}
\usepackage{cite}

\newcommand{\up}{\vspace{0cm}}

\title{Do players reason by forward induction\\ in dynamic perfect information games?}
\author{Sujata Ghosh
\institute{Indian Statistical Institute\\ Chennai, India}
\email{sujata@isichennai.res.in}
\and
Aviad Heifetz
\institute{The Open University of Israel\\ Raanana, Israel}
\email{aviadhe@openu.ac.il}
\and
Rineke Verbrugge
\institute{University of Groningen\\ Groningen, The Netherlands}
\email{rineke@ai.rug.nl}
}
\def\titlerunning{Do players reason by forward induction in dynamic perfect information games?}
\def\authorrunning{Sujata Ghosh, Aviad Heifetz \& Rineke Verbrugge}

\begin{document}


\maketitle


\begin{abstract}
We conducted an experiment where participants played a perfect-information game
against a computer, which was programmed to deviate often from its
backward induction strategy right at the beginning of the game. Participants
knew that in each game, the computer was nevertheless optimizing against some belief about
the participant's future strategy. 

It turned out that in the aggregate,  participants were likely to respond in a way which is optimal with respect to their best-rationalization extensive form rationalizability conjecture - namely the conjecture that the computer is after a larger prize than the one it has foregone, even when this necessarily meant that the computer has attributed future irrationality to the participant when the computer made the first move in the game. Thus, it appeared that participants applied forward induction.
However, there exist alternative explanations for the choi\-ces of most participants;  for example, choices could be based on the extent of risk aversion that participants attributed to the computer in the remainder of the game, rather than to the sunk outside option that the computer has already foregone at the beginning of the game.  
For this reason, the results of the experiment do not yet provide conclusive evidence for Forward Induction reasoning on the part of the participants. 
\end{abstract}






\section{Introduction}\label{sec:intro}


Backward Induction (BI) is a canonical approach for solving extensive-form
games with perfect information. In generic games with no payoff ties, BI
yields the unique subgame perfect equilibrium~\cite{perea12,heifetz12}. Nevertheless, BI embodies a
conceptual difficulty: in subgames following a deviation of some player (or
players) from their BI\ strategy, it is not obvious why players should
necessarily believe that the deviators will `return to their senses' and
realign their behavior in the subgame with the BI\ dictum. Because such certainty is absent, BI\ might itself be suboptimal for players who are skeptical
about such re-adherence to rationality. Thus, the epistemic assumptions underpinning BI are
those of relentlessly reborn optimism (see e.g. the surveys in \cite{perea07} 
and \cite{perea14}), with all players, under all contingencies, commonly
believing in everybody's future rationality, no matter how irrational
players' past behavior has already proven: \textquotedblleft after all,
tomorrow is another day!\textquotedblright 

\begin{figure*}[t]
\begin{center}
\includegraphics[width=.98\textwidth]{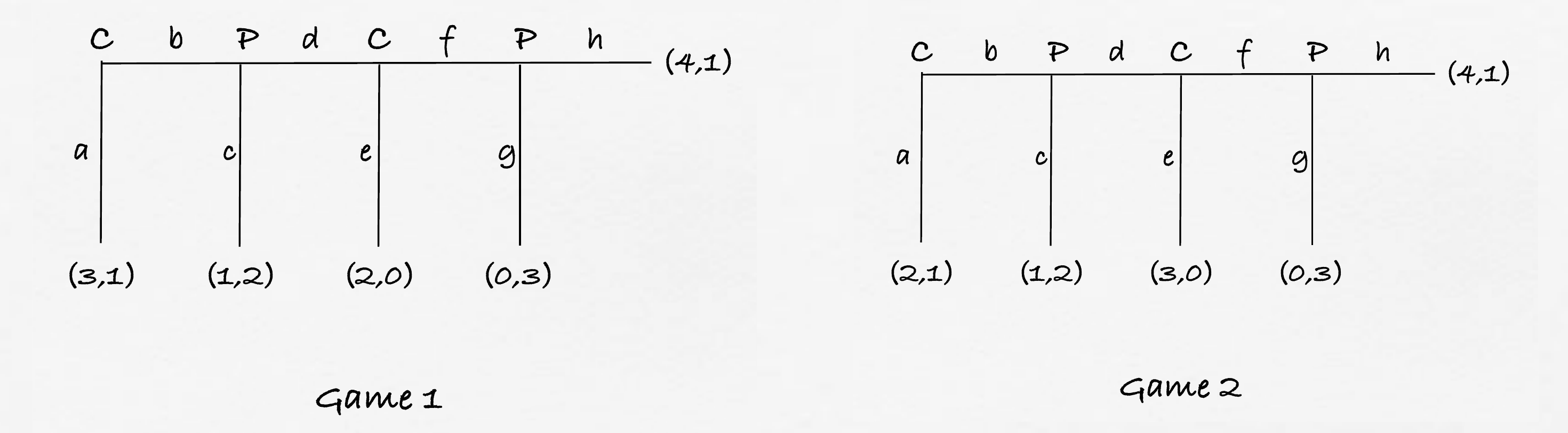} 
\includegraphics[width=.98\textwidth]{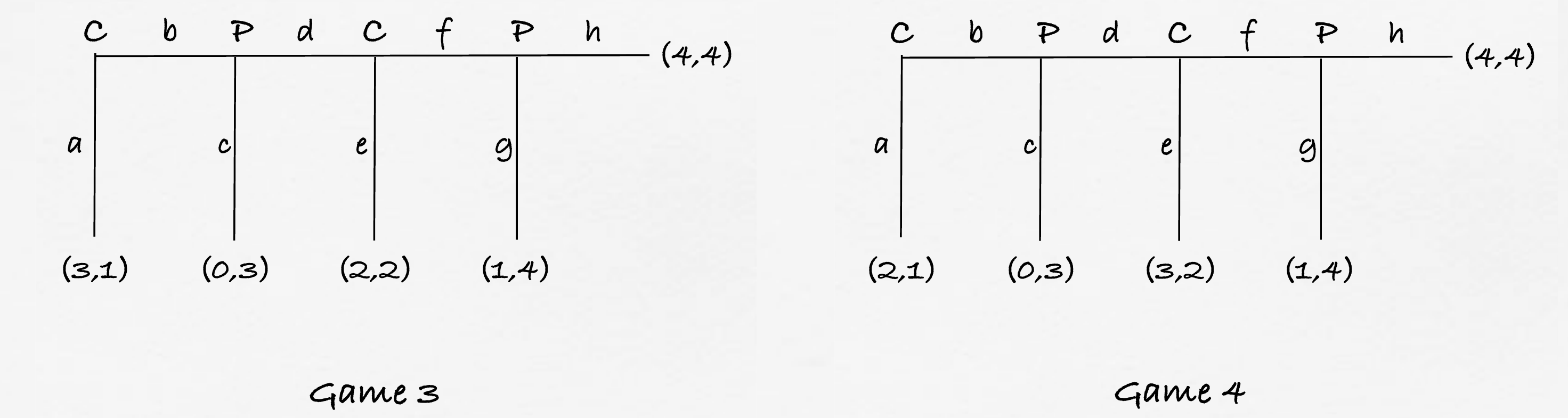} 
\end{center}
\caption[]{Collection of the main games used in the experiment. The ordered pairs at the leaves represent pay-offs for the computer ($C$) and the participant ($P$), respectively.} 
\label{fig:maingames}
\end{figure*}

An alternative, more sober approach on the part of a player may be to employ
Forward Induction (FI) reasoning, and to try to rationalize her opponent's past
behavior in order to assess his future moves. For example, even in a subgame
where there exists no strategy of the opponent which is consistent with common
knowledge of rationality \emph{and }his past behavior, she may still be able
to rationalize his past behavior by attributing to him a strategy which is
optimal as against a presumed \emph{suboptimal} strategy of hers. Or, even
better, it may sometimes be possible for her to attribute to him a strategy
which is optimal with respect to a \emph{rational} strategy of hers, which is,
though, in return only optimal as against a suboptimal strategy of \emph{his}%
. If the player pursues this rationalizing reasoning to the highest extent
possible \cite{battigalli96} and reacts accordingly, she will end up choosing
an Extensive-Form Rationalizable (EFR) strategy \cite{pearce84,battigalli97}.

EFR strategies may thus be distinct from BI\ strategies, as an example by Reny
\cite{reny92} shows (see game 1 in Figure~\ref{fig:maingames}). Given this difference, it is
therefore a completely non-trivial result that in perfect information games
with no relevant ties,\footnote{%
That is, where each player has a strict ranking over all the game-tree leaves
following each of her decision nodes.} there is nevertheless a unique EFR\ 
\emph{outcome}, which coincides with the unique BI outcome \cite{battigalli97,chenmicali11,chenmicali13,perea12,hp14}. Only when
relevant payoff ties are allowed, an outcome-discrepancy between the two
solution concepts may appear. In such cases, the EFR outcomes constitute a
subset of the BI\ outcomes \cite{chenmicali11,chenmicali13,perea12}, and the
inclusion may be strict, as demonstrated by Chen and Micali \cite{chenmicali13} (see game
3 in Figure~\ref{fig:maingames}).

We note here that experimental studies in behavioral economics have shown that the backward induction outcome is often not reached in large centipede games. Instead of immediately taking the `down' option, people often show partial cooperation, moving right for several moves before eventually choosing `down'~\cite{mckelvey1992,nagel98,camerer03}. Nagel and Tang \cite{nagel98} suggest that people sometimes have reason to believe that their opponent could be an altruist who usually cooperates by moving to the right and McKelvey and Palfrey \cite{mckelvey1992} suggest that players may believe that there is some possibility that their opponent has payoffs different from the ones the experimenter tries to induce by the design of the game.

We could also ask the following question: {\em Are people inclined to use forward induction when they play a game, and
in particular in games like those of Reny or Chen-Micali mentioned above?} This
question was the motivation for the experiment on which we report here. Our
pivotal interest was to examine participants' behavior following a deviation from
BI behavior by their opponent right at the beginning of the game. 

\begin{figure*}[t]
\begin{center}
\includegraphics[width=.98\textwidth]{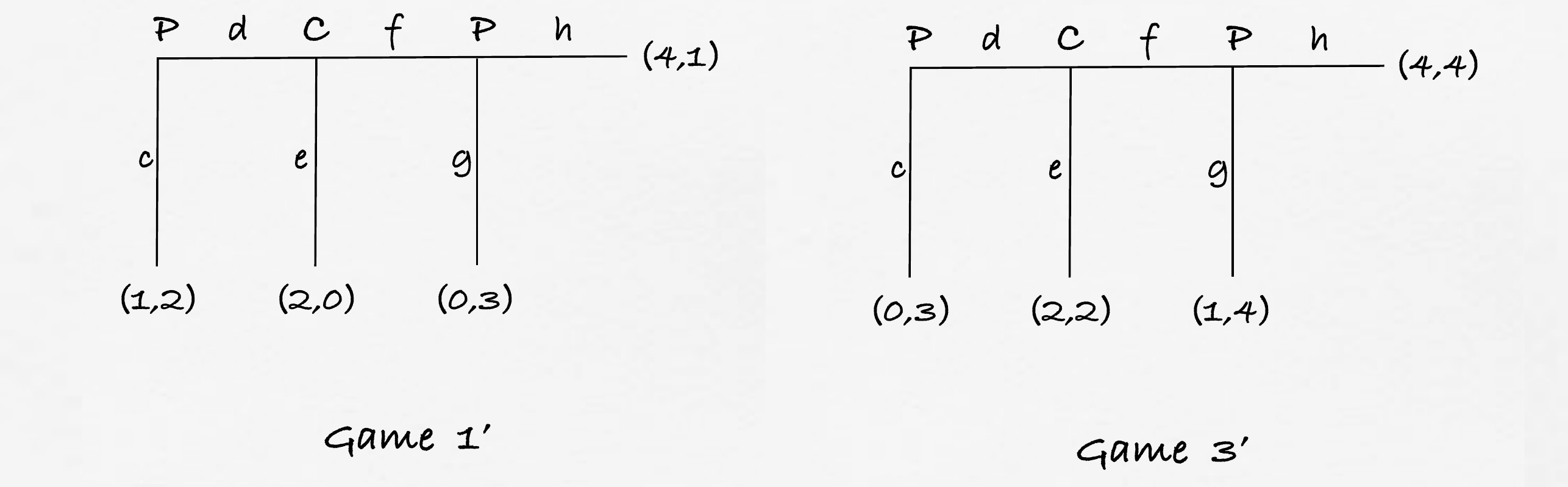} \up\up\up
\end{center}
\caption[]{Truncated versions of Game 1 and Game 3. The ordered pairs at the leaves represent pay-offs for $C$ and $P$, respectively.\up}
\label{fig:auxgames}
\end{figure*}

\subsection{Designing an experiment about forward induction behavior}

When designing an experiment to tackle the question whether people are inclined to use FI when they play dynamic perfect-information games, 
 such as the Reny or Chen-Micali games mentioned  above, a first challenge was
to neutralize repeated-game effects across repetitions of the same game. 
 Such visible repetitions could  enable a folk-theorem style augmented cooperation level
among participants playing one against the other, bypassing their cooperation
opportunities in a one-shot play of the same game (cf.~\cite{fudenberg1986}). We chose to address this
challenge by letting participants, adult university students with little or no knowledge of game theory, knowingly play against a computer.\footnote{Another important advantage of using computer opponents in experiments with dynamic games is that the experimenter can control the strategies used by the computer opponent, which allows better interpretation of the participants' decisions. Using a computer opponent also has disadvantages, for example, players might reason quite differently about their opponent if they know they are playing against a human player. Interestingly, Hedden and Zhang~\cite{hedden2002} misinformed a part of their subjects that they were playing against a human opponent while in fact everyone was playing against a computer. They found little difference between the decisions of these groups, and only around 10 $\%$ of participants expressed a
suspicion on an exit questionnaire that they were playing against a computer rather than a person.}

 We programmed the
computer so as to follow, in each repetition of each game, a strategy which
is optimal with respect to some strategy of the human participant. This strategy for
the computer was decided in advance for each round, so that the computer did
not learn from experience in previous games. This information was honestly and simply conveyed
to the participants at the beginning of the experiment, by the following item on the
instruction sheet (see Appendix A):

\begin{quote}
How does the computer reason in each particular game of the experiment?

- The computer thinks that you already have a plan for that game, and it plays
the best response to the plan it thinks that you have for that game.  

- However, the computer does not learn from previous games and does not take
into account your choices during the previous games.
\end{quote}

Given that the participants were playing against a computer, a second challenge
was to create variability in the appearance   of repetitions of the
same game, so that each repetition looks different and forces the participant to
think anew about her or his strategy in the current repetition. This was
achieved by two measures:

 \begin{itemize}
\item[-] repeating in each round a set of  6 games, distinct in terms of pay-off structures (see more
on the games in Section \ref{sec:games}); and 
 \item[-] presenting the game to the
participant in a different graphical fashion in each round in which the game was 
repeated. The game play was animated, and proceeded by
consecutive dropping of a marble through trapdoors controlled alternately by
the players, leading ultimately to one of several possible bins with orange
marbles for the participant and blue marbles for the computer.\footnote{The game presentation was inspired by Ben Meijering's `marble drop' games, also used in~\cite{ghosh2014,meijering2011,Meijering2014}.} In repetitions of
the same extensive-form game, changes were made in right/left directions of
trapdoors in junctions on paths leading to each bin (see, for example,  
screenshots of different representations of the same game in Appendix C, Figure \ref{fig:experimentgames}).
\end{itemize}

In the earlier experiments that investigated FI reasoning in human participants, the experimental games mostly considered an outside option together with some form of imperfect information games, see, e.g.,  \cite{bn08,shahriar14,cachon1996,huck2005}. Such games are more complicated in nature than the dynamical games of perfect information we consider here. The novelty of the current experiment lies in its simplicity, using perfect-information games only.


\section{Experimental games}\label{sec:games} 


The list of games that were used in the experiment is given in Figure \ref{fig:maingames} and Figure \ref{fig:auxgames}. In these two-player games, the players play alternately. Let $C$ denote the computer, and let $P$ denote the participant. 
In the first four games (Figure  \ref{fig:maingames}), the computer plays first, followed by the participant, and each of the players can play at two decision nodes. In the last two games (Figure \ref{fig:auxgames}), which are truncated versions of two of the games in Figure~\ref{fig:maingames}, the participant gets first chance to move. We will now discuss the BI and EFR (FI) strategies of all the 6  games; these are summarized in Table 1.

\subsection{BI and EFR strategies in four main games}

Game 1 has been introduced by Reny \cite{reny92}. Here, the unique Backward Induction (BI) strategies for player $C$ and player $P$ are $a;e$ and $c;g$, respectively. In case the last decision node of the game is reached, player $P$ will play $g$ (which will give $P$ better payoff at that node) yielding 0 for $C$. Thus, in the previous node, if reached, $C$ will play $e$ to be better off. Continuing like this from the end to the start of the game (by BI reasoning) it can be inferred that whoever is the current player will play so as to end the game immediately. 

Forward induction, in contrast, would proceed as follows. Among the two strategies of player $C$ which are compatible with reaching the first decision node of player $P$, namely $b; e$ and $b;f$, only the latter is rational for player $C$. This is because of the fact that $b;e$ is dominated by $a;e$, while $b;f$ is optimal for player $C$ if she believes that player $P$ will play $d;h$ with a high enough probability. Attributing to player $C$ the strategy $b;f$ is thus player $P$'s best way to rationalize player $C$'s choice of $b$, and in reply, $d;g$ is player $P$'s best response to $b;f$. Thus, the unique Extensive-Form Rationalizable (EFR) strategy of player $P$ is $d;g$, which is distinct from her BI strategy $c;g$. Nevertheless, player $C$'s best response to $d;g$ is $a;e$, which is therefore player $C$'s EFR strategy. Hence the EFR outcome of the game (with the EFR strategies $a;e$, $d;g$) is identical to the BI outcome. This is an instance of the general theorem~\cite{battigalli97,chenmicali13,aa12,hp14} mentioned in the Introduction, by which in perfect-information games with no relevant payoff ties, the unique BI outcome coincides with the unique EFR outcome (even when, as in this game, for some player the EFR strategy is different from the BI strategy).

Game 2 is   popularly known as the Centipede game~\cite{rosenthal81}. Here, the structure of the tree is as in the Reny game (cf. Figure 1, game 1), but the payoffs of player $C$ following $a$ and $e$ are interchanged. As in game 1,   the unique Backward Induction (BI) strategies of player $C$ and player $P$ are also $a;e$ and $c;g$, respectively (and the BI outcome is the leaf following $a$). However, when considering FI reasoning for game 2, unlike in game 1, there does exist a belief of player $C$ with respect to which $b;e$ is optimal. This is the belief that player $P$ is playing with   high probability the strategy $d;g$, a strategy that is actually optimal for player $P$ if $P$ believes that $C$ is playing with   high probability $b;f$, which in turn is optimal for player $C$ if $C$ believes that player $P$ is playing with   high probability $d;h$ and is thus irrational only at the last decision node.  For this reason, in game 2 it turns out that $a;e$ and $c;g$ are the unique EFR strategies of the corresponding players, and hence coincide with their unique BI strategies. 

Game 3 has been introduced by Chen and Micali \cite{chenmicali13}. Note that player $P$ has identical payoffs at both leaves following her second and final decision node. As a result, there are two ways to fold the game backwards, and every action of every player at each decision node is a Backward Induction (BI) choice. Consequently, all possible outcomes of the game are BI outcomes. However, the strategy $b;e$ of player $C$ is dominated by its strategy $a;e$; and thus $b;e$ is not an Extensive-Form Rationalizable (EFR) strategy for player $C$. In contrast, $b;f$ is optimal for player $C$ under the belief that player $P$ will pursue $d;h$ with a high enough probability. Hence, if player $P$ finds herself in her first decision node, her best way to rationalize player $C$'s first move of $b$ is to attribute to it the strategy $b;f$, to which only $d;g$ and $d;h$ are best replies. Therefore, the path $b;c$ with the eventual payoffs (0, 3) for players $C$ and $P$, respectively, is not an EFR outcome of the game. This is an instance of the general result by~\cite{chenmicali11,chenmicali13} mentioned in the Introduction, by which the set of EFR outcomes is a (possibly proper) subset of the set of BI outcomes.

Finally, in game 4, the structure of the tree is as in the Chen and Micali game (cf. Figure \ref{fig:maingames}, game 3), but the payoffs of player $C$ following $a$ and $e$ are interchanged with respect to game 3. Here too, every action of every player at each decision node is a BI choice, and hence all possible outcomes of the game are BI outcomes. However, in this case, each of the three strategies that player $C$ has -- namely strategy $a$, strategy $b;e$ and strategy $b;f$ -- is a best reply to some conjecture about player $P$'s strategy. Similarly to the game 2 scenario, $b;e$ is a best reply to the conjecture that player $P$ is likely playing $d;g$, and $b;f$ is a best reply to the conjecture that player $P$ is likely playing $d;h$. Thus, for player $P$, in case her first decision node is reached, both her choices $c$ and $d$ constitute rationalizable (EFR) choices. Hence, in this case, all possible outcomes are EFR outcomes as well, identical to the BI outcomes.

\subsection{BI and EFR strategies in truncated games}

 Game $1'$ is a truncated version of game 1, with player $P$ being the starting player. The BI strategy for player $P$ in this game is to play $c$. In case player $P$ plays $d$ and the first decision node of player $C$ is reached, both BI and EFR strategies for player $C$ are the same $-$ to play $e$. Thus the EFR strategy for player $P$ in this game is to play $c$, and the BI and EFR outcomes coincide, as they should in finite perfect-information games without relevant ties.

Game $3'$ is a truncated version of game 3, with player $P$ starting the game. Player $P$ has identical payoffs at the leaves following her second decision node. Here again, every action of every player at each decision node is a BI choice, and hence all possible outcomes of the game are BI outcomes. In case the first decision node of player $C$ is reached, each of the two strategies that player $C$ has -- namely strategy $e$ and strategy $f$ -- is a best reply to some conjecture about player $P$'s strategy. Strategy $e$ is a best reply to the conjecture that player $P$ is likely playing $d;g$, and $f$ is a best reply to the conjecture that player $P$ is likely playing $d;h$.  Thus, for player $C$, in case its second decision node is reached, both its choices $e$ and $f$ constitute rationalizable (EFR) choices. Hence, in this case also, all possible outcomes are EFR outcomes as well, identical to the BI outcomes, in contrast to what happens in game 3. 

\begin{table}
\begin{center}
\begin{tabular}{ || l | l | l || }
\hline\hline
{\small\bf Games | Strategies} & {\small BI strategy} & {\small EFR strategy} \\ 
\hline
{\small Game 1} & {\small C: $a;e$} & {\small C: $a;e$} \\
				    & {\small P: $c;g$} & {\small P: $d;g$} \\	
\hline
{\small Game 2} & {\small C: $a;e$} & {\small C: $a;e$} \\
				    & {\small P: $c;g$} & {\small P: $c;g$} \\
\hline
{\small Game 3} & {\small C: $a;e, b;e, a;f, b;f$} & {\small C: $a;e, a;f, b;f$} \\
				    & {\small P: $c;g, d;g, c;h, d;h$} & {\small P: $d;g, d;h$} \\	
\hline
{\small Game 4} & {\small C: $a;e, b;e, a;f, b;f$} & {\small C: $a;e, b;e, a;f, b;f$} \\
				    & {\small P: $c;g, d;g, c;h, d;h$} & {\small P: $c;g, d;g, c;h, d;h$} \\
\hline
{\small Game $1'$} & {\small C: $e$} & {\small C: $e$} \\
				    & {\small P: $c;g$} & {\small P: $c;g$} \\	
\hline
{\small Game $3'$} & {\small C: $e, f$} & {\small C: $e, f$} \\
				    & {\small P: $c;g, d;g, c;h, d;h$} & {\small P: $c;g, d;g, c;h, d;h$} \\
\hline\hline
\end{tabular}\up\up
\label{game-summary}
\end{center}
\caption{BI and EFR (FI) strategies for the 6 experimental games in Figures 1 and 2}
\end{table}


\section{Experimental procedure}\label{sec:design}

\begin{figure}[h]
\begin{center}
\includegraphics[width=.48\textwidth]{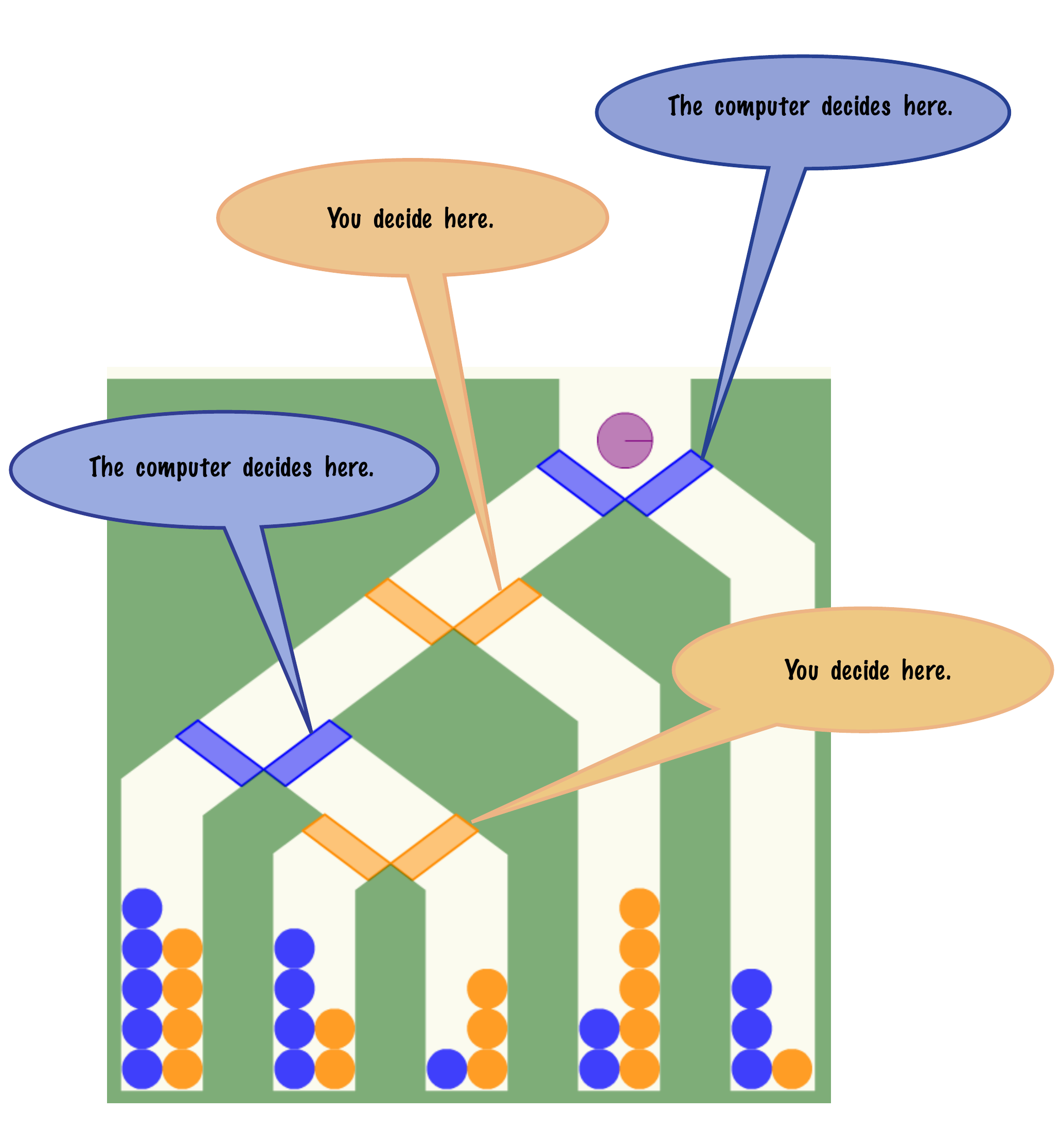} \up\up\up\up\up
\end{center}
\caption[]{Graphical interface for the participants. The computer controls the blue trapdoors and acquires pay-offs in the form of blue marbles (represented as dark grey in a black and white print), while the participant controls the orange trapdoors and acquires pay-offs in the form of orange marbles (light grey in a black and white print).} 
\label{fig:interface}
\end{figure}

The experiment was conducted at the Institute of Artificial Intelligence (ALICE) at the University of Groningen, The Netherlands.  
A group of 50 Bachelor's and Master's students from different disciplines at the university took part in this experiment. The participants had little or no knowledge of game theory, so as to ensure that neither backward induction nor forward induction reasoning was already known to them. The participants played the finite perfect-information games in a graphical interface on the computer screen (cf. Figure \ref{fig:interface}).  In each case, the opponent was the computer, which had been programmed to play according to plans that were best responses to some plan of the participant. The participants were instructed accordingly. In each game, a marble was about to drop, and both the participant and the computer determined its path by controlling the orange and the blue trapdoors: The participant controlled the orange trapdoors, and the computer controlled the blue trapdoors. The participant's goal was that the marble should drop into the bin with as many orange marbles as possible. The computer's goal was that the marble should drop into the bin with as many blue marbles as possible.  



 At first, 14 practice games were played (see Figure~\ref{fig:practice}, Appendix C), which were simpler than the 6 games outlined in Section~\ref{sec:games}. At the end of each practice game, the participant could see how many marbles he or she had gained in that game, and also the total number of marbles  gained so far. These games were presented in increasing levels of difficulty in terms of the reasoning the participants needed to perform with respect to their and the opponent's (computer's) choices, to maximize their gains. 


The 14 practice games were followed by 48 experimental games and the participants got access to similar information regarding the number of marbles gained. There were 8 rounds, each comprised of the 6 games that were described in Section \ref{sec:games}. Different graphical representations of the same game were used in different rounds (cf. Figure \ref{fig:experimentgames}, Appendix C). A break of 5 minutes was given after the participants  finished playing 4 rounds of the experimental games.
 The participants earned between 10 and 15 euros for participating in the experiment. The amount depended on the number of marbles won during the experimental phase, and they were told about this before the start of the experiment. They earned 10 euros for participation, and each marble a participant won added 4 cents to the amount. The final amount was rounded off to the nearest 5 cents mark.
 
\begin{figure}[h]
\begin{center}
\includegraphics[width=.45\textwidth]{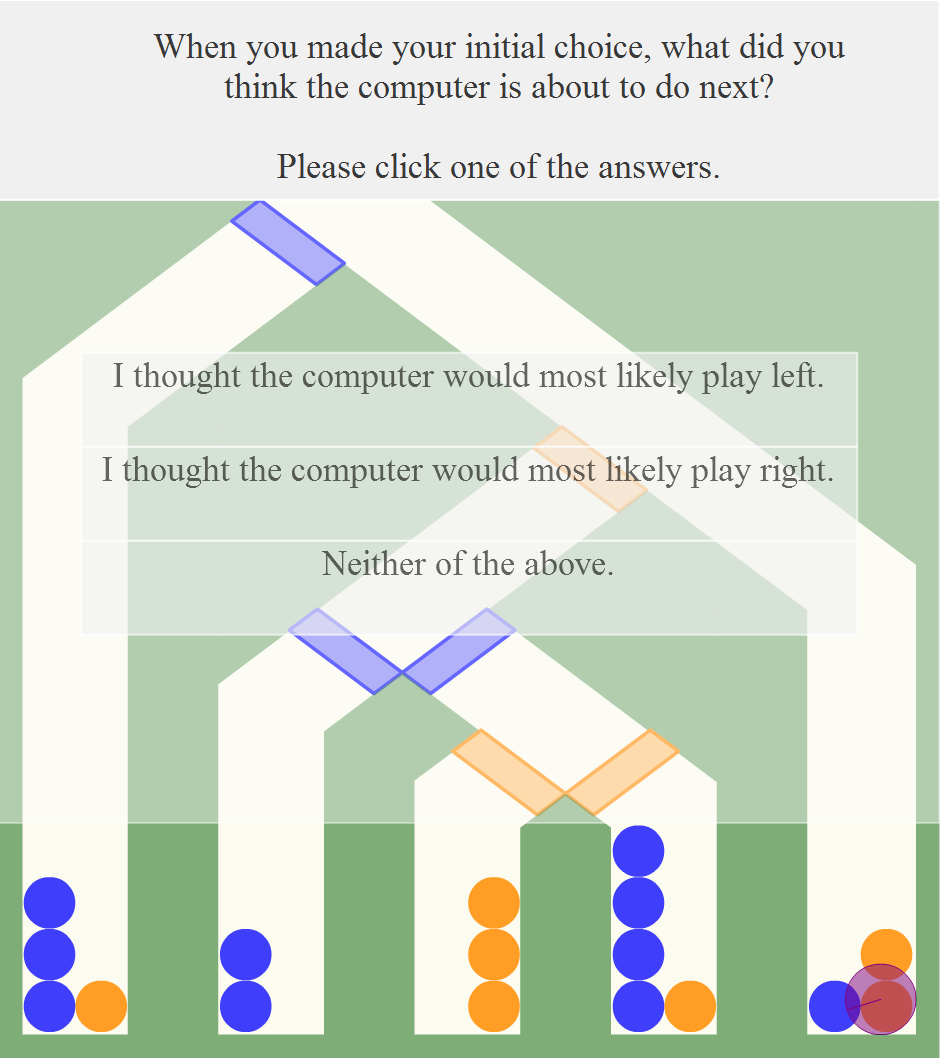} \up\up\up\up
\end{center}
\caption[]{Question on computer's behavior} 
\label{fig:question}
\end{figure}

At some points during the experimental phase, the participants were asked a multiple-choice question as follows: ``When you made your initial choice, what did you think the computer was about to do next?'' (cf. Figure \ref{fig:question}). Three options were given regarding the likely choice of the computer: ``I thought the computer would most likely play left" or ``I thought the computer would most likely play right" or ``neither of the above".  The first two answers translated to the moves $e$ or $f$ of the computer, respectively. In case of the third answer, we assumed that  the participant was undecided regarding the computer's next choice. 
The participants had been randomly divided into two groups: Group A and Group B, each consisting of 25 persons. The members of group A were asked the question about the computer's next possible move once they had played at their first decision node in each game in rounds 3, 4, 7, and 8, whereas the members of group B were asked the same question but less often, namely only in each game in rounds 7 and 8.

At the end of the experiment, each participant was asked the following question: ``When you made your choices in these games, what did you think about the ways the computer would move when it was about to play next?'' The participant needed  to describe   the plan he or she thought was followed by the computer on its next move after the participant's initial choice, in his or her own words. In summary, during the experiment, the participants had to perform the tasks specified in Table 2 in the order given there. 

\begin{table}
\begin{center}
\begin{tabular}{ || l | l ||}
\hline\hline
{\small Step 1} & {\small Introduction and instructions.}\\
\hline
{\small Step 2} & {\small Practice Phase: 14 games.}\\
\hline
{\small Step 3} & {\small - Experimental Phase: 48 game items, divided into }\\
           & {\small 8 rounds of 6 different games each, in terms of  }\\
           & {\small isomorphism class of pay-off structures;}\\
           & {\small - Each of the 6 games occurs once in each round;  }\\
           & {\small these games occur in the same order in each round;} \\
           & {\small - Question on computer's behavior (cf. Figure \ref{fig:question}) in }\\
           & {\small several rounds: Group A in rounds 3, 4, 7, 8; Group B }\\
           & {\small in rounds 7, 8.}\\
\hline
{\small Step 4} & {\small Final Question.}\\
\hline\hline 
\end{tabular}\up\up
\label{expt-summary}
\end{center}
\caption{Steps of the experiment}
\end{table}

During the 48 games of the experimental phase, played by each participant, a varied amount of data were collected. In particular, for each participant, for each game, for each round of the game, we collected the following data:
\begin{itemize} 
\item[-] Participant's decision at his/her first decision node, if the node was reached. In particular, whether move $c$ or $d$ had been played.\footnote{In addition, we also took note of other aspects, such as  the participant's behavior at the second decision node and time taken by the participant at various stages. We leave out the details, because these are not relevant for our main research question, whether participants are applying forward induction. See Appendix B  
for recorded data types, and see~\cite{halder2015} for a typology of players's reasoning strategies based on these richer data.}

\end{itemize}


\section{Results and analysis}\label{sec:results}


As mentioned above,   
we report and analyze only the behavior of the
participants in their first decision node, that is their choice between actions $c$ or $d$
whenever that decision node was reached. We found no significant variation (Proportion test, p = 0.21) between the behavior of the 25
participants of Group A, who were asked questions (cf. Figure \ref{fig:question}) after each game in rounds
3, 4, 7, and 8, and the 25 participants of Group B, who were asked those
questions only after the games in rounds 7 and 8. Therefore henceforth, we will analyze the data of
all 50 participants together.

\begin{figure*}[t]
\begin{center}
\begin{tabular}{cc}
\includegraphics[width=.5\textwidth]{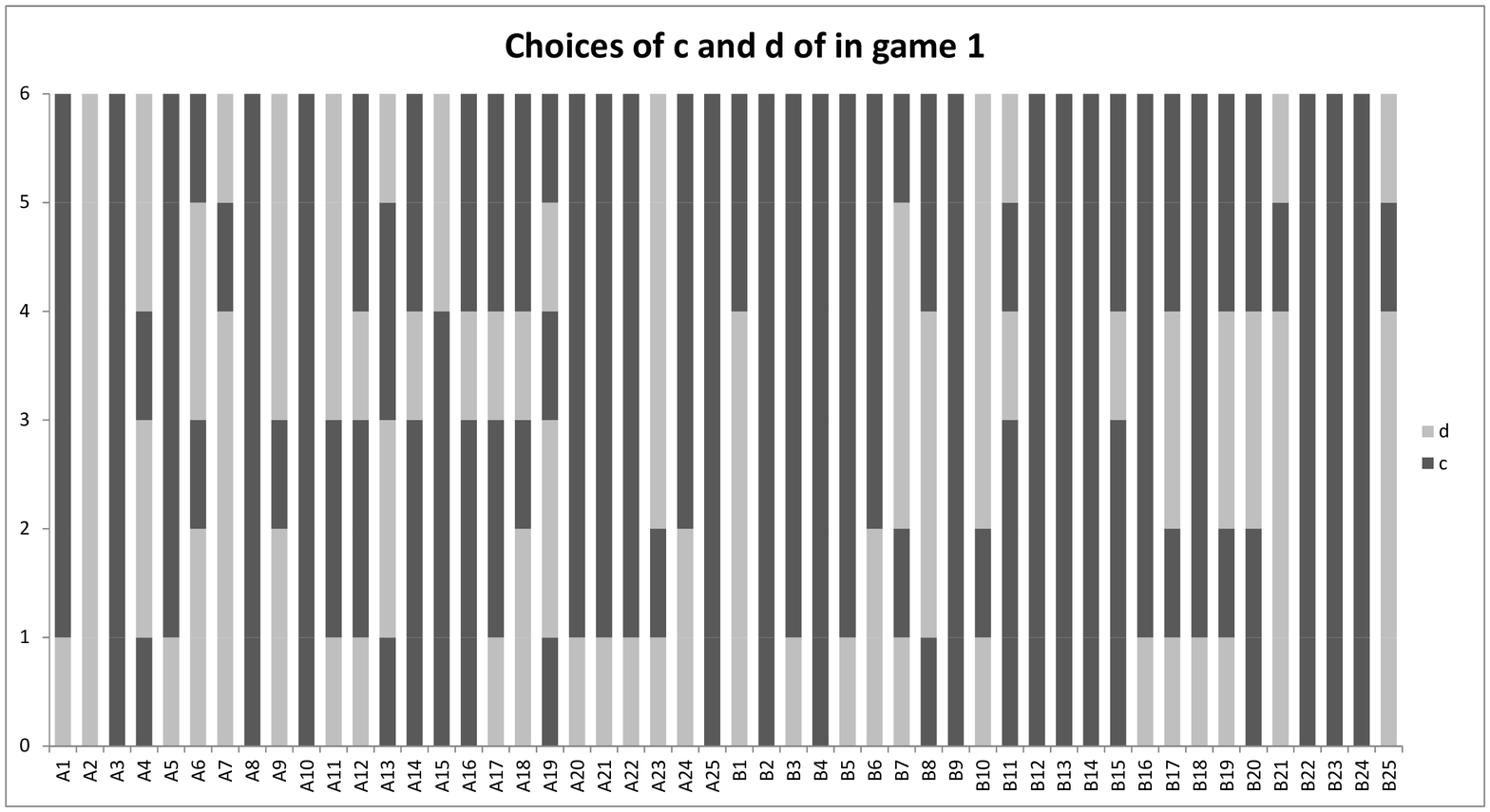} & \includegraphics[width=.5\textwidth]{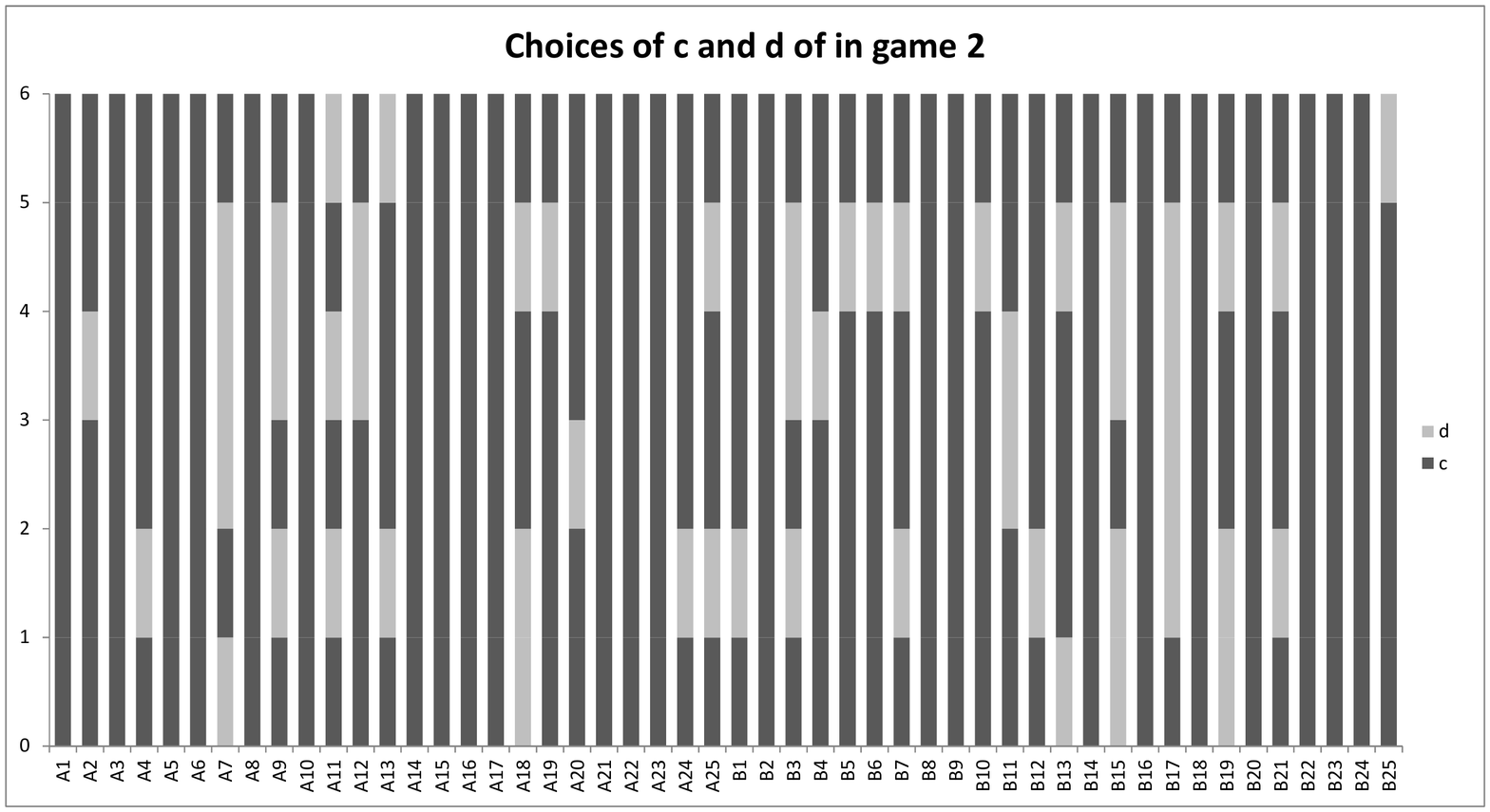}\up\up\up\up\up  \up\up\up\up\\ 
\includegraphics[width=.5\textwidth]{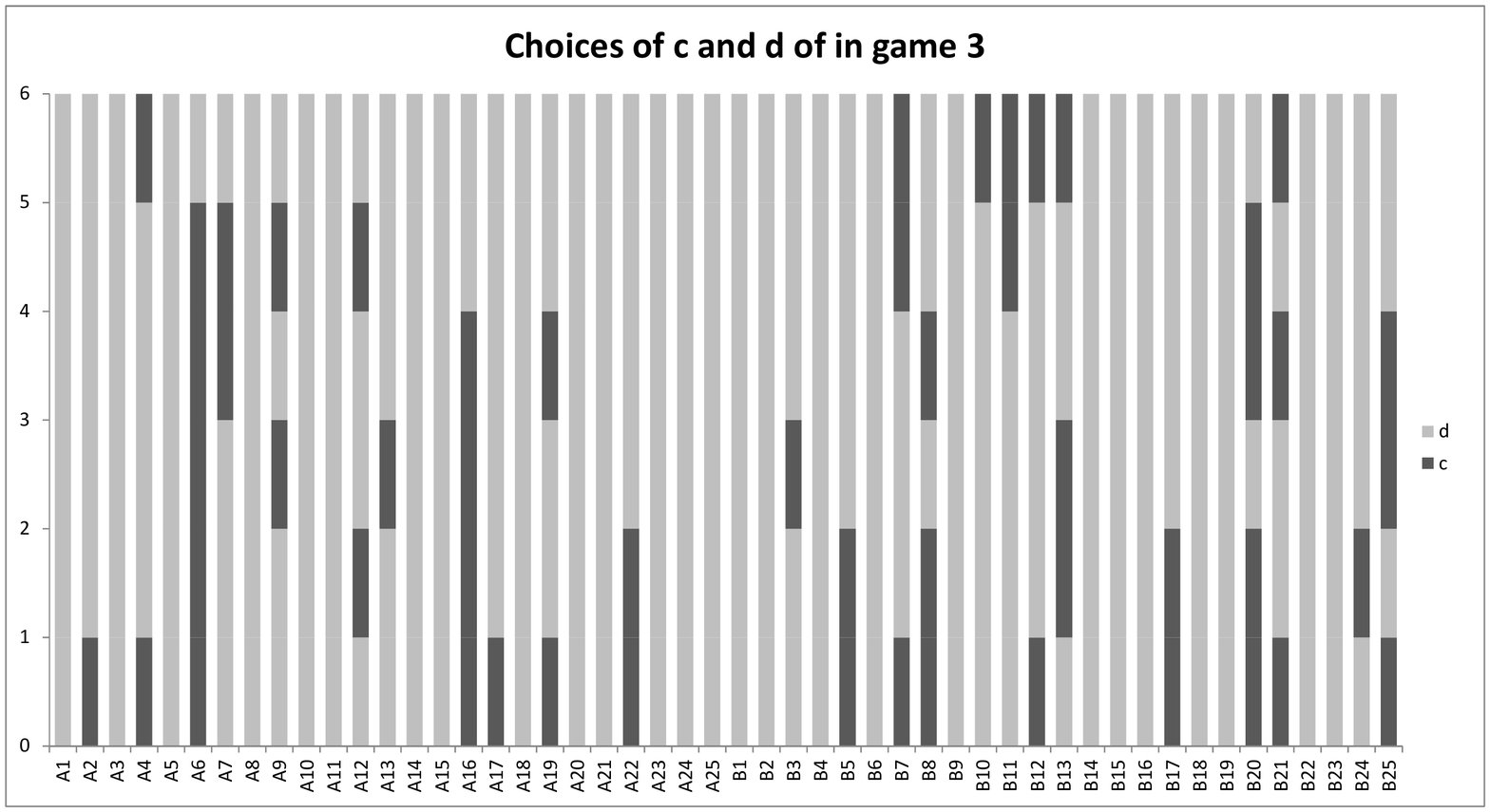} & \includegraphics[width=.5\textwidth]{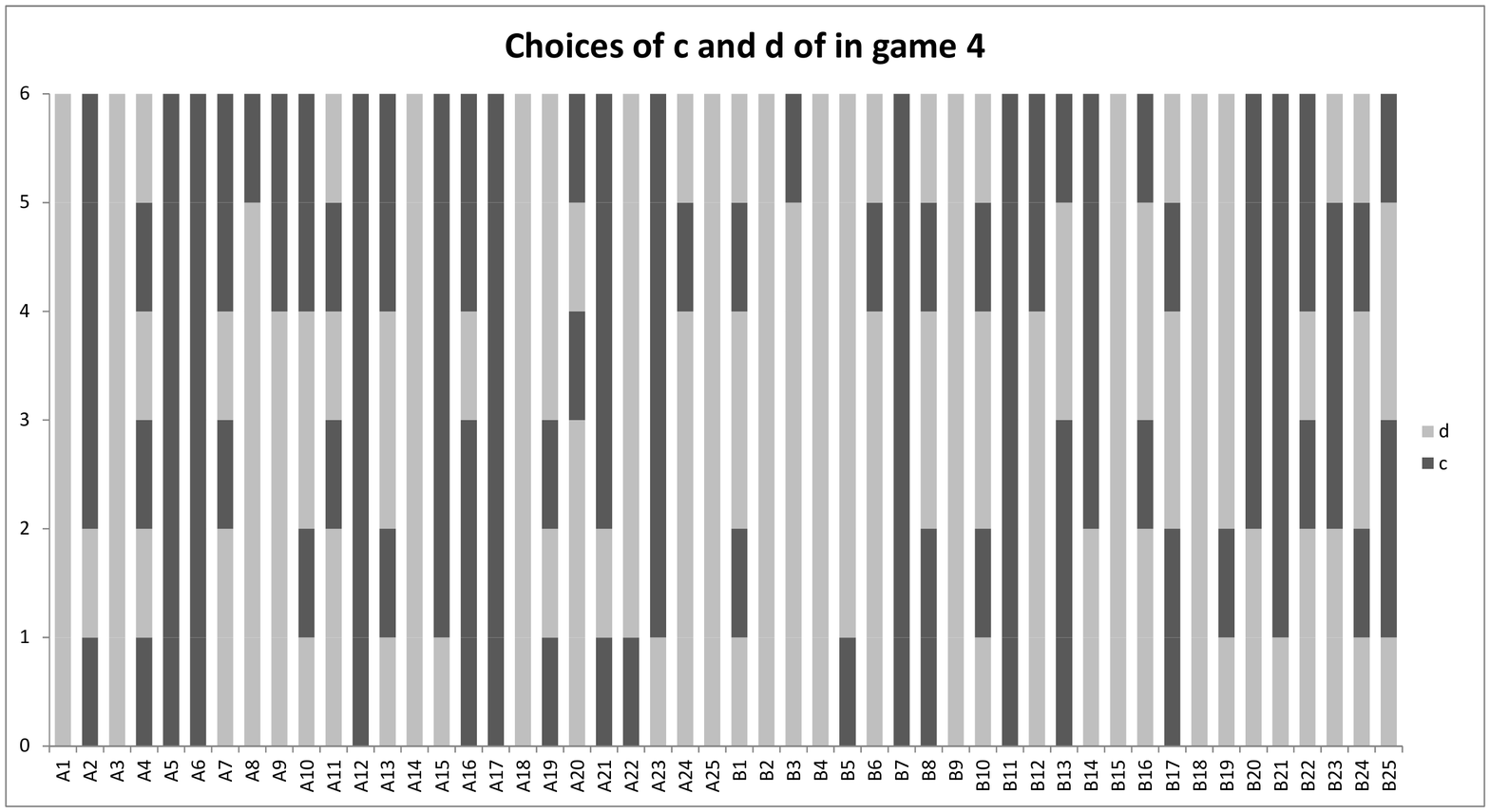}\up\up\up\up\up  \up\up\up\up\\ 
\end{tabular}
\end{center}
\caption[]{Sequence of choices (across the 8 repetitions of each game) at the first decision node of games 1, 2, 3, 4, per participant (named A1 \ldots A25, B1  \ldots B25). The {\em dark grey} color corresponds to the rounds the participant played the move $c$, and the {\em light grey} color corresponds to the rounds the participant played move $d$, whenever the participant's first decision node was reached.\up\up\up} 
\label{fig:choices1}
\end{figure*}

\begin{figure*}[t]
\begin{center}
\begin{tabular}{cc}
\includegraphics[width=.5\textwidth]{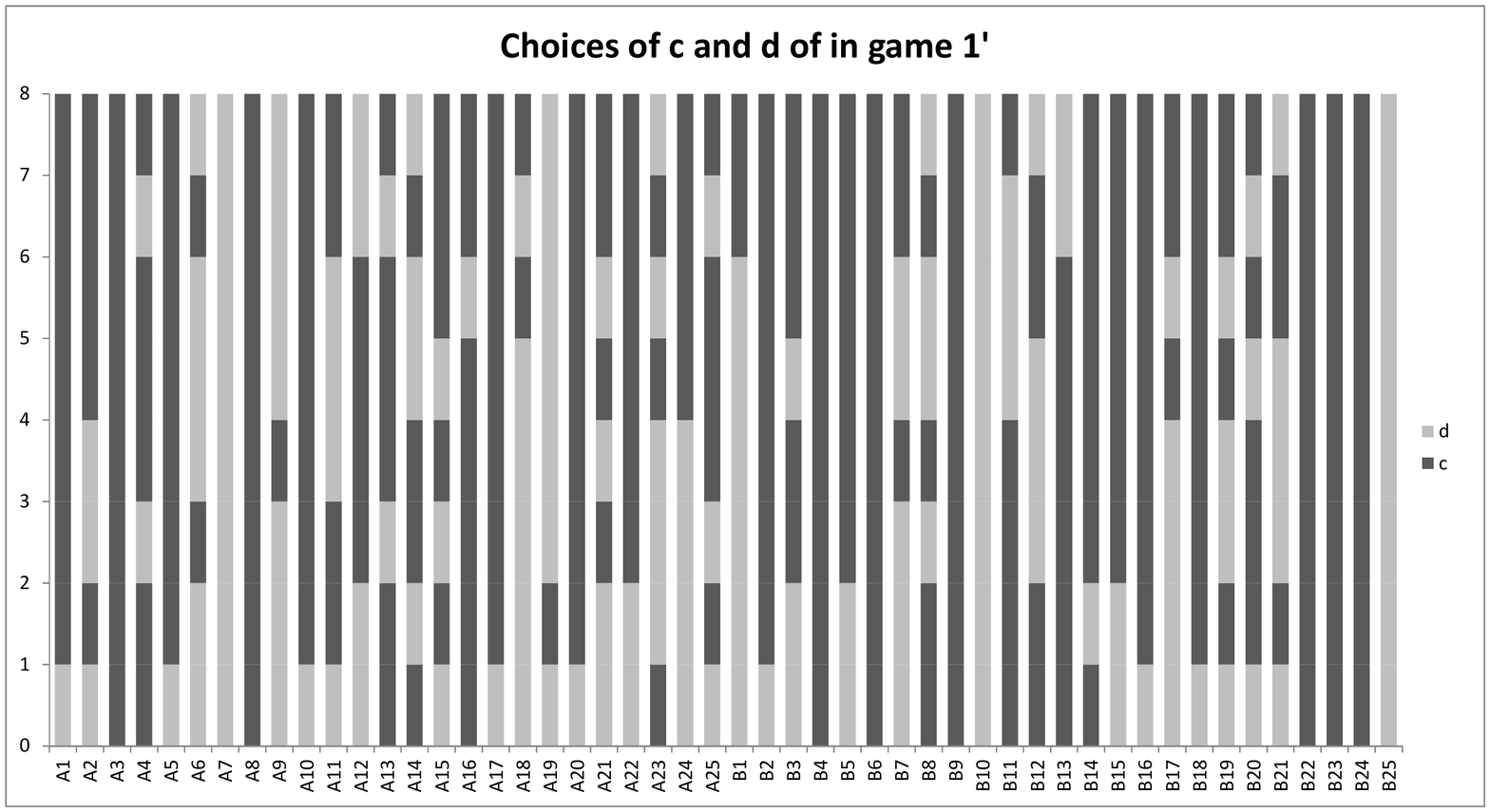} & \includegraphics[width=.5\textwidth]{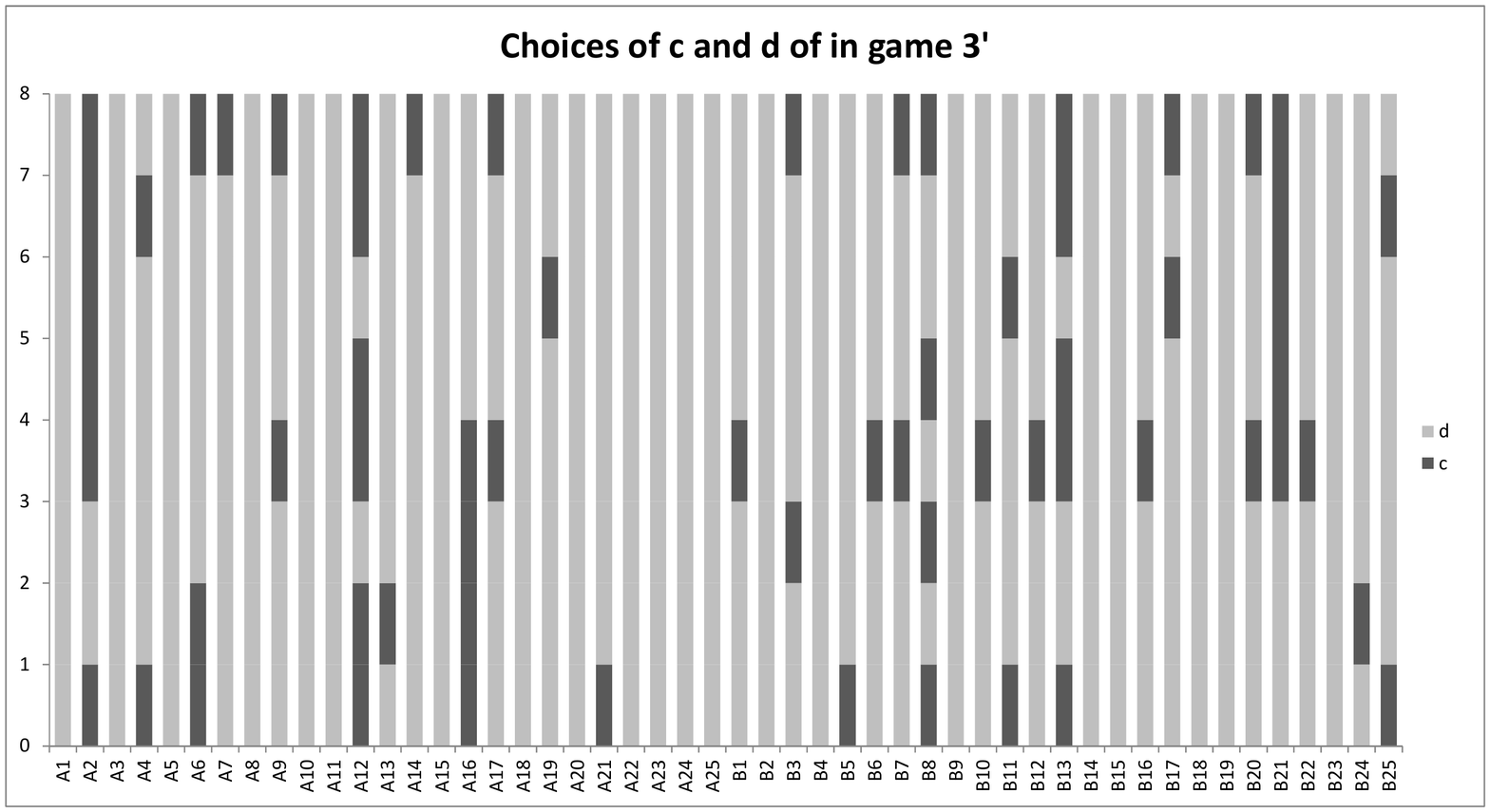}\up \up \up\up\up  \up\up\up\up\\ 
\end{tabular}
\end{center}
\caption[]{Sequence of choices (across the 8 repetitions of each game) at the first decision node of games $1', 3'$, per participant (named A1 \ldots A25, B1  \ldots B25). The {\em dark grey} color corresponds to the rounds the participant played the move $c$, and the {\em light grey} color corresponds to the rounds the participant played move $d$, whenever the participant's first decision node was reached.\up\up\up} 
\label{fig:choices2}
\end{figure*}

The four graphs of Figure~\ref{fig:choices1} and the two graphs of Figure~\ref{fig:choices2} give the sequence of choices (across the repetitions of each game) at the first decision node, per participant (named A1 \ldots A25, B1  \ldots B25), for games 1, 2, 3, 4 and for games $1', 3'$, respectively. The {\em dark grey} color corresponds to the rounds the participant played the move $c$, and the {\em light grey} color corresponds to the rounds the participant played move $d$, whenever the participant's first decision node was reached. They clearly show that $d$ was played more often in game 1 than in game 2
(which has the same payoffs as game 1 except for $C$'s payoffs interchanged at
two leaves). Moreover, $d$ was played more often in game 3 than in game 4 (which similarly has the
same payoffs as game 3 except for $C$'s payoffs interchanged at two leaves). These
observations may initially suggest corroboration of FI reasoning because (as the reader can check in Table 1), $d$ is $P$'s only
EFR move in game 1 while $c$ is the only EFR move in game 2, and $d$ is the only
EFR move in game 3 while both $c$ and $d$ are EFR moves in game 4. This would provide a positive answer to our research question whether players apply forward induction when playing against a computer which sometimes deviates from rational behavior.



However, a closer look at individual choices while also taking the truncated games $1'$ and $3'$ of Figure 2 into account, casts doubt that these
findings can be attributed to any substantial FI reasoning. 
When comparing game 1 to game $1'$, EFR prescribes $d$ in game 1 and $c$ in
game $1'$ (see Table 1). However, only two participants out of 50 (4\%) played $d$ much more often in game 1 than in game $1'$%
;\footnote{%
The verbal elaboration of one of the two participants at the end of the experiment is indeed
compatible with EFR, see Appendix D.} four additional participants (8\%) played $d$ in game 1 only slightly more often than
in game $1'$; but 24 other participants (48\%) actually played $d$ more often in $1'$
than in 1!   

Similarly, when comparing game 3 to game $3'$, EFR prescribes $d$ in game 3,
while in game $3'$, both $c$ and $d$ are compatible with EFR. However, only two
participants out of 50 (4\%) played $d$ much more often in game 3 than in
game $3'$; ten additional participants (20\%) played $d$ in game 3
only slightly more often than in game $3'$; but 17 other participants (34\%)
actually played $d$ more often in game $3'$ than in game 3. In summary, comparing games 1 and 3 to their truncated versions does not lend support for FI reasoning.

Now, comparing game 3 to game 4, we find that 47 participants (94\%) played $d$ at least as
often in game 3 as in game 4, and the remaining three players (6\%) played $d$ only slightly less often in game 3 than in
game 4. As mentioned at the beginning of this section, at first glance this may then
suggest support for EFR behavior (since EFR prescribes $d$ in game 3 and allows
for both $c$ and $d$ in game 4). However, because we did not see support for EFR when
comparing game 3 to game $3'$, it could very well be that a cardinal effect
rather than an ordinal effect has played a role here: 

\newpage

\begin{itemize}
\item[-] In game 4,  a participant's playing $d$ implied that the computer would
have to choose between a payoff 3 that it could reach for certain by going down, and a `lottery' between the
payoffs 1 and 4 that it would meet if it would continue to the right to the next $P$-node, due to the fact that at $P$'s last decision node, the participant $P$ gains the same payoff of 4 points after either choice. Consequently, most participants might have
feared that the computer would go for the certain payoff 3, so preempted
that by choosing $c$. 

\item[-] In game 3, in contrast, a participant's playing $d$ implied that the computer would
have to choose between the relatively low payoff 2 that it could achieve for sure by going down, and again a `lottery' between the
payoffs 1 and 4. Consequently, most participants may have been betting that the
computer would go for the `lottery', and hence chose $d$.\footnote{%
Some verbal comments at the end attributed to the computer a 50\%-50\%
belief in this lottery and expected payoff maximization, which is indeed consistent
with choosing  $c$ in game 4 and $d$ in game 3.} 
\end{itemize}

\newpage

Similarly, comparing game 1 to game 2, we find that 42 out of 50 participants (84\%) played
$d$ at least as often in game 1 as in game 2.
Here again, at first glance this may seem to lend support for EFR behavior,
since EFR prescribes $d$ in game 1 and $c$ in game 2. However, here too, a
cardinal effect may have played a role, as follows: 

\begin{itemize}
\item[-] In game 2, a payoff of 3 may have seemed (in the eyes of most
participants) to tempt the computer to go down at its second decision point and settle for it for sure, rather than
hoping that the participant would err at the end by choosing $h$ -- an error which would yield only a
slightly better payoff of 4 to the computer, while $C$'s pay-off would be only 0  if the participant did not err at the end.

\item[-] In game 1, in contrast, participants may have attributed a greater
temptation to the computer to gamble for the payoff 4 (which is what the
computer would get if the participant were to err by choosing $h$) versus 1 (if the participant  did not
err); the participant would compare this `lottery' with what $C$ could settle for with certainty by going down at its second decision point, which is only 2. 
\end{itemize}

These considerations may have led most players to choose $d$ more often in
game 1 than in game 2, irrespective of any FI considerations.


\section{Conclusion}\label{sec:concl}


To the best of our knowledge, the experiment carried out and reported here is the first experiment that has been  designed to test Forward Induction (FI) behavior (particularly, Extensive-Form Rationalizable (EFR) behavior) in extensive-form games with {\bf perfect information}. 
 
In the experiment, 50 participants played against a computer, which they knew to have been programmed so as not to make deductions or learn from previous game rounds, but rather to optimize, in each round, against some belief about the participant's strategy. Moreover, different rounds of the same game were interspersed  in between different rounds of other games, and in different rounds of the same game the game tree was presented to the participants in distinct interactive ``marble-drop" forms. Thus, unlike in other experiments where each pair of participants plays repeatedly many rounds of the same game, our design was structured so as to neutralize, as much as possible, repeated-game cooperation considerations on the part of each participant.  

In the aggregate, the participants were more likely to respond in a way which is optimal with respect to their best-rationalization EFR conjecture - namely the conjecture that the computer is after a larger prize than the one it has foregone, even when this necessarily meant that the computer has attributed future irrationality to the participant when the computer made the first move in the game. Thus, it appeared that participants did apply forward induction.

However, there exist alternative explanations for the choi\-ces of most participants, and such alternative explanations also emerge from several of the participants' free-text verbal descriptions of their considerations (cf. Appendix D), as solicited from them at the end of the experiment. These alternative considerations have to do with the extent of risk aversion that participants attributed to the computer in the remainder of the game, rather than to the sunk outside option that the computer has already foregone at the beginning of the game.  
For this reason, the results of the experiment do not yet provide conclusive evidence for Forward Induction reasoning on the part of the participants. 

In current ongoing work, we are using data from this experiment, such as response times and answers to questions, in order to investigate how participants can be divided into meaningful classes according to other cognitive considerations, for example, whether they are applying quick, instinctive thinking or contemplative, slower deliberation, whether they are applying higher orders of theory of mind, and so on, see~\cite{halder2015}. In future work, we aim to investigate  which strategies participants actually apply in dynamic games with perfect information in which the opponent occasionally deviates from backward induction.  We plan to use new games with different pay-off structures and will perform an eye-tracking study to check the points in the games to which participants attend while reasoning.

\section*{Acknowledgements}

We would like to acknowledge research assistants Sourit Man\-na, Damian Podareanu, and Michiel van de Steeg, without whose help this work would not have been possible. We would also like to thank Harmen de Weerd for his contribution to the graphs in this article.  Furthermore, we are grateful to Shaul Tsionit for his statistical advice. Sujata Ghosh and Rineke Verbrugge acknowledge the Netherlands Organisation of Scientific Research for Vici grant NWO 227-80-001 awarded to the latter.

\bibliographystyle{eptcs}
\bibliography{experiment2}






\newpage
\section*{Appendix A: Instruction sheet}\label{sec:sheet}

\begin{itemize}
\item[-] In this task, you will be playing two-player games. The computer is the other player.
\item[-] In each game, a marble is about to drop, and both you and the computer determine its path by controlling the orange and the blue trapdoors.
\item[-] You control the orange trapdoors, and the computer controls the blue trapdoors.
\item[-] Your goal is that the marble drops into the bin with as many orange marbles as possible. The computer's goal is that the marble drops into the bin with as many blue marbles as possible.
\item[-] Click on the left trapdoor if you want the marble to go left, and on the right trapdoor if you want the marble to go right.
\item[-]How does the computer reason in each particular game?
\begin{quote}
The computer thinks that you already have a plan for that game, and it plays the best response to the plan it thinks that you have for that game.

However, the computer does not learn from previous games and does not take into account your choices during the previous games.
\end{quote}
\item[-] The first 14 games are practice games. At the end of each practice game, you will see how many marbles you gained in that game, and also the total number of marbles you have gained so far.
\item[-] The practice games are followed by 48 experiment games. At the beginning of the experiment games, the total number of marbles won will be set at 0 again. At the end of each experiment game, you will see how many marbles you gained in that game, and also the total number of marbles you have gained so far.
\item[-] You will be able to start each game by clicking on the ``START GAME'' button, and move to the next game by clicking on the ``NEXT'' button.
\item[-] At some points during the experiment phase, you will be asked a few questions regarding what guided your choices.
\item[-] There will be a break of 5 minutes once you finish 24 of the 48 experiment games.
\item[-] The money you will earn is between 10 and 15 euros and depends on how many marbles you have won during the experiment phase. You will get 10 euros for participation, and each marble you win will add 4 cents to your amount. The final amount will be given to you rounded off to the nearest 5 cents mark.\footnote{We chose the relatively large `show-up fee' because Dutch student participants tend to complain in case of large differences in pay between participants. However, most participants attained a fairly large award, so in future we aim to incentivise participants more by offering a lower show-up fee and a higher fee per marble.}

\end{itemize}


\section*{Appendix B: Recorded data types}\label{sec:data}



As mentioned earlier, 50 students participated in this experiment. The participants were first requested to provide the following information:

\begin{quote}
\hspace{.2cm} Name; Age; Gender; Field of study.
\end{quote}

 Then they were given instruction sheets mentioning what they were supposed to do (see Appendix A) 
 together with a representative figure (cf. Figure \ref{fig:interface}) of the graphical interface of the games they were supposed to play. Once they got accustomed with what they were expected to do, 
the participants played the first 14 practice games.  As mentioned in Section \ref{sec:design}, at the end of each game, a participant could see how many orange marbles he or she had won till that moment - this was to show how his/her winnings were getting calculated. At the end of the practice phase, the experimental phase began.

Here, each participant played 48 experimental games, playing each of the six games depicted in Figures \ref{fig:maingames} and \ref{fig:auxgames}, eight times, in different representations. During these 48 games, played by each participant, a varied amount of data were collected. 
For each participant, for each game, for each round of the game, we collected the following data:

\begin{itemize} 
\item[-] participant's decision at his/her first decision node, if the node was reached. In particular, whether move $c$ or $d$ had been played;
\item[-] participant's decision at his/her second decision node, if the node was reached. In particular, whether move $g$ or $h$ had been played; 
\item[-] time taken by the participant in starting the game, i.e. the time between the moment the game was shown to the participant, and the moment he/she clicked on the ``start'' button; 
\item[-] time taken by the participant in making his/her decision at the first decision node, if the node was reached, i.e. the time between the moment the computer passed the playing marble to the participant on its first decision node, and the moment he/she clicked on the next trapdoor for the marble to be dropped;
\item[-] time taken by the participant in making his/her decision at the second decision node, if the node was reached, i.e. the time between the moment the computer passed the playing marble to the participant on its second decision node, and the moment he/she clicked on the next trapdoor for the marble to be dropped.
\end{itemize}

\begin{figure}[h]
\begin{center}
\includegraphics[width=.38\textwidth]{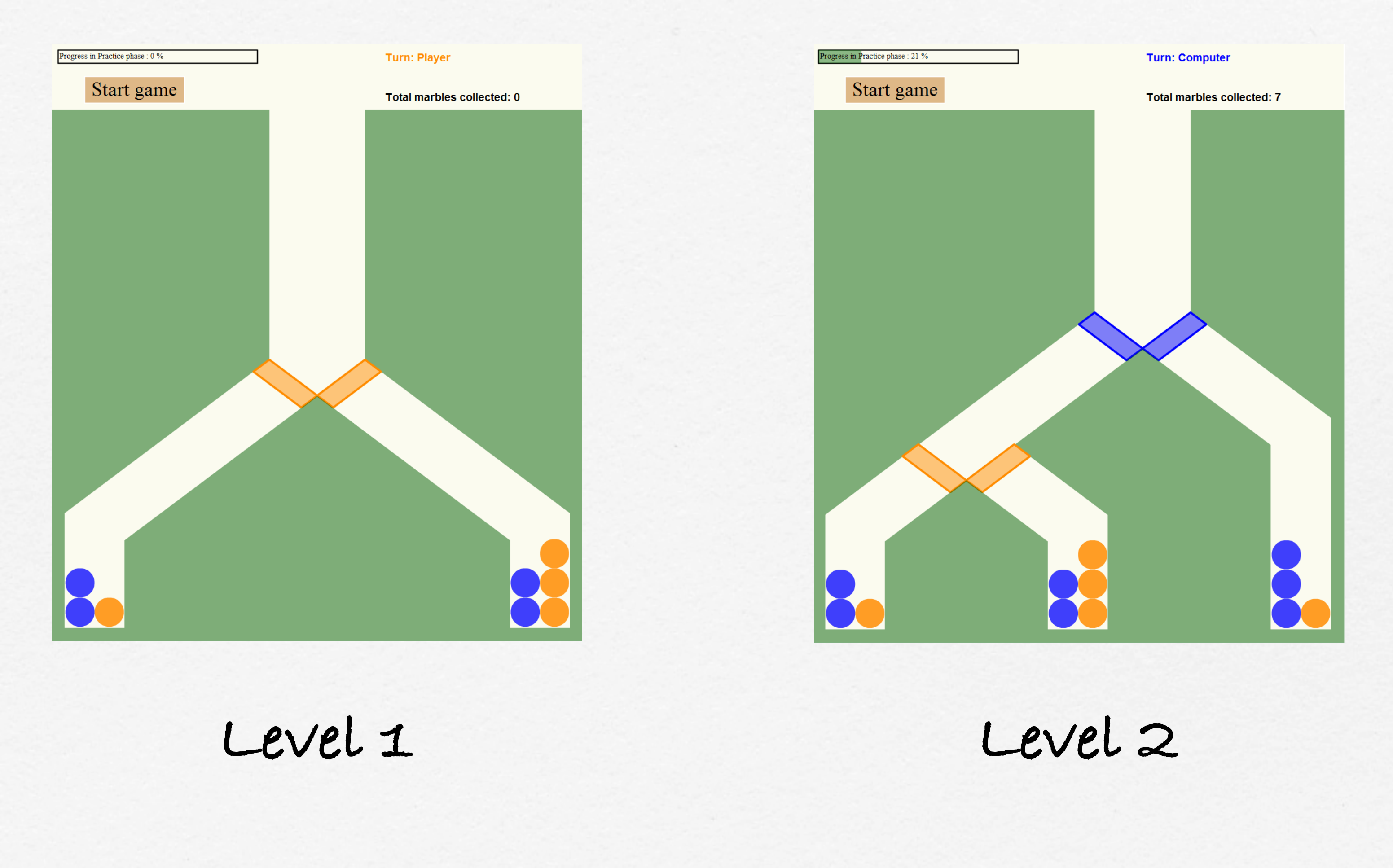} \\
\includegraphics[width=.38\textwidth]{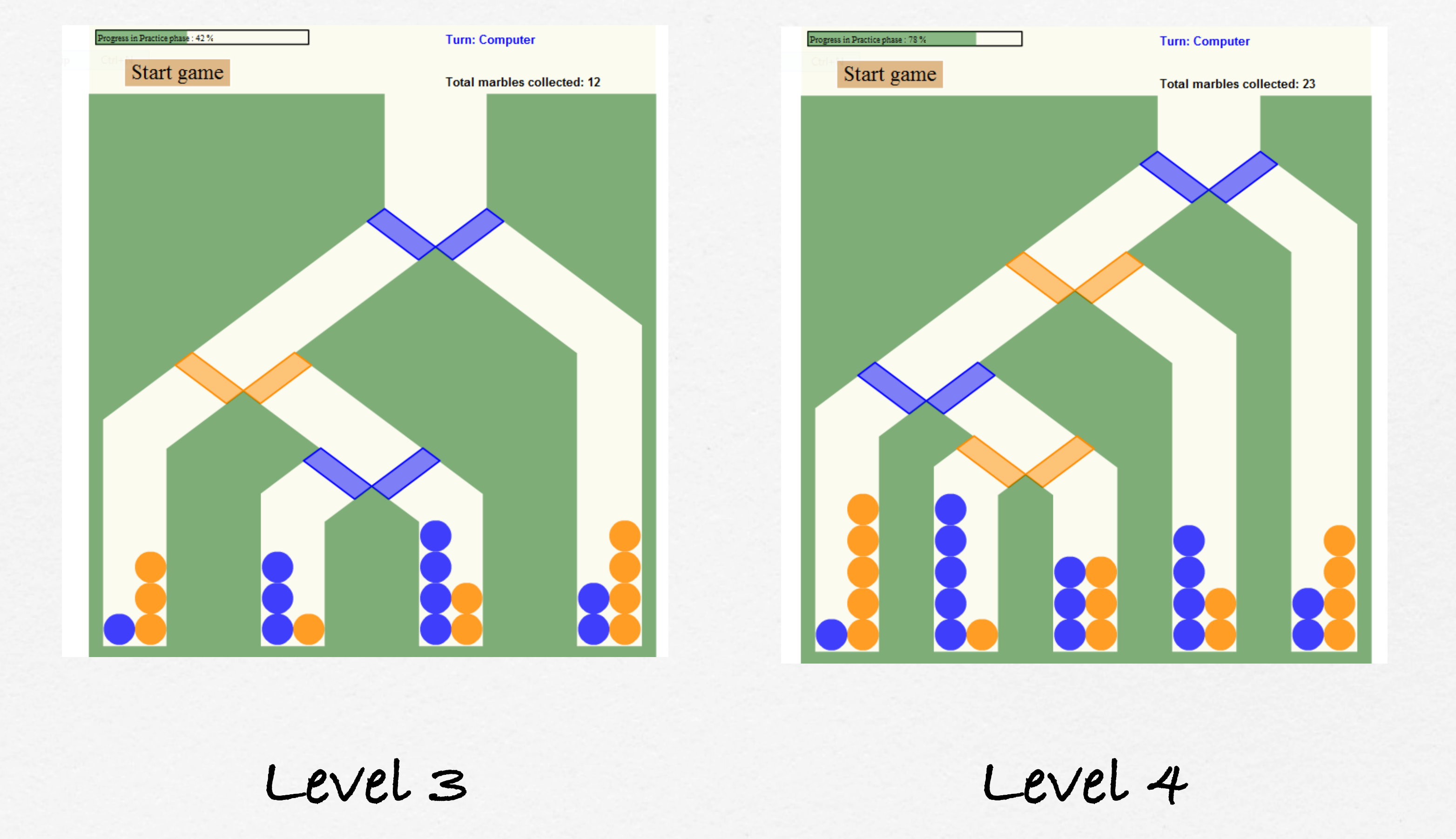} \up \up \\ 
\end{center}
\caption[]{Levels of practice games} 
\label{fig:practice}
\end{figure}

The first two items correspond to categorical or qualitative data, whereas the next three, which are  response times recorded in milliseconds, correspond to numerical or quantitative data. As mentioned in Section \ref{sec:design}, the participants were randomly divided into two groups. namely Group A and Group B, where members of group A were asked to answer a question (cf. Section \ref{sec:design}) in rounds 3, 4, 7, 8, and members of group B were asked to answer the same questions but only in rounds 7 and 8. 
For each participant,   depending on the group  (Group A or Group B),
we collected the following data:

\begin{itemize}
\item[-] participant's answer to the given question (cf. Figure \ref{fig:question}) at the ends of the rounds in which it was asked. In particular, whether the answer was $e$ or $f$ or undecided;
\item[-] time taken by the participant in giving the answer, i.e. the time between the moment the question appeared on the screen and the moment he/she clicked on his/her choice of answer.
\end{itemize}

 The first data item is categorical, whereas the second one, recorded in milliseconds, is numerical. Finally, at the end of the experiment each participant was asked a final question (cf. Section \ref{sec:design}), 
the answers to which were recorded in a separate sheet. 
A limited amount of space was given  in which the answer was  to be formulated. 

\section*{Appendix C: Experimental interface}\label{sec:interface}

During the training phase, the participants were given 14 training games of increasingly difficult levels in terms of number of decision points, as explained in Section~\ref{sec:design}. Figure~\ref{fig:practice} shows example games for each of the four levels.

\begin{figure}[ht]
\begin{center}
\begin{tabular}{ll}
\includegraphics[width=.2\textwidth]{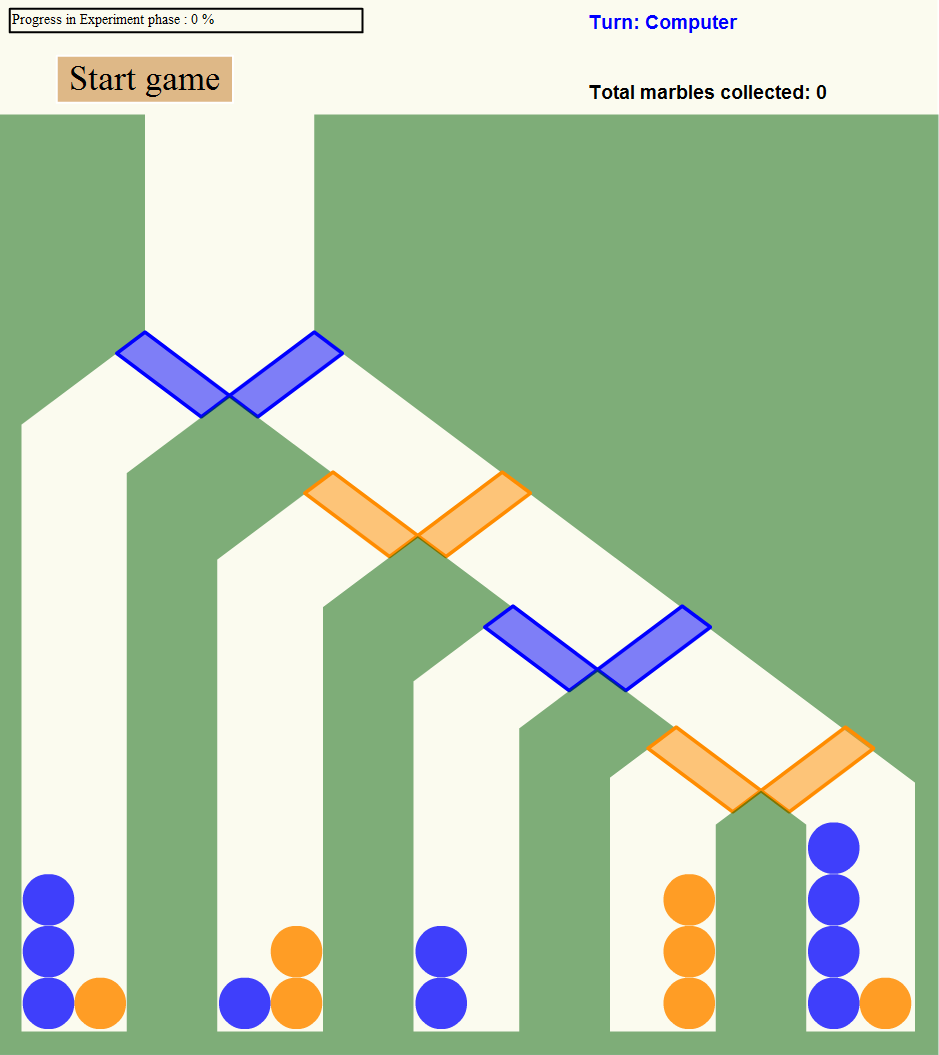} & \quad\includegraphics[width=.22\textwidth]{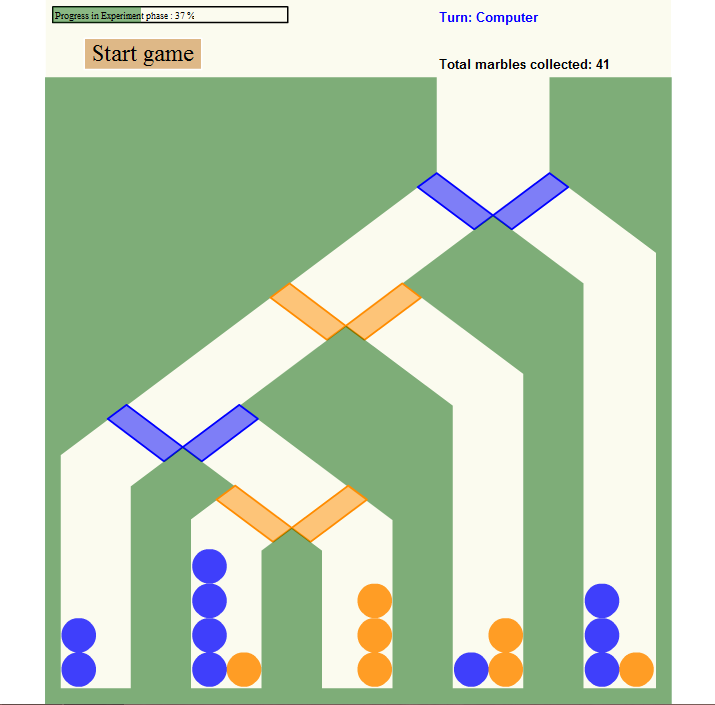}\\
& \\
\includegraphics[width=.22\textwidth]{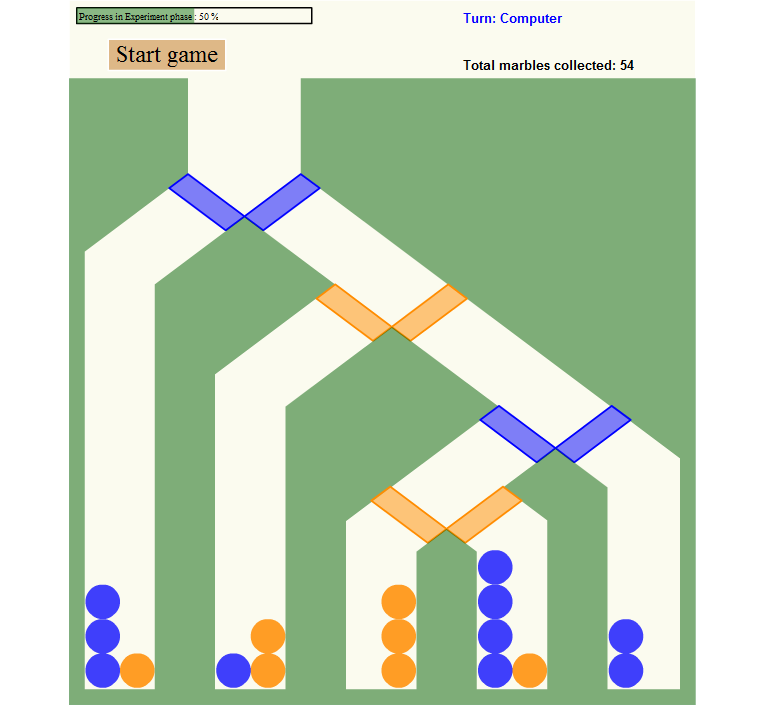} & \quad\includegraphics[width=.22\textwidth]{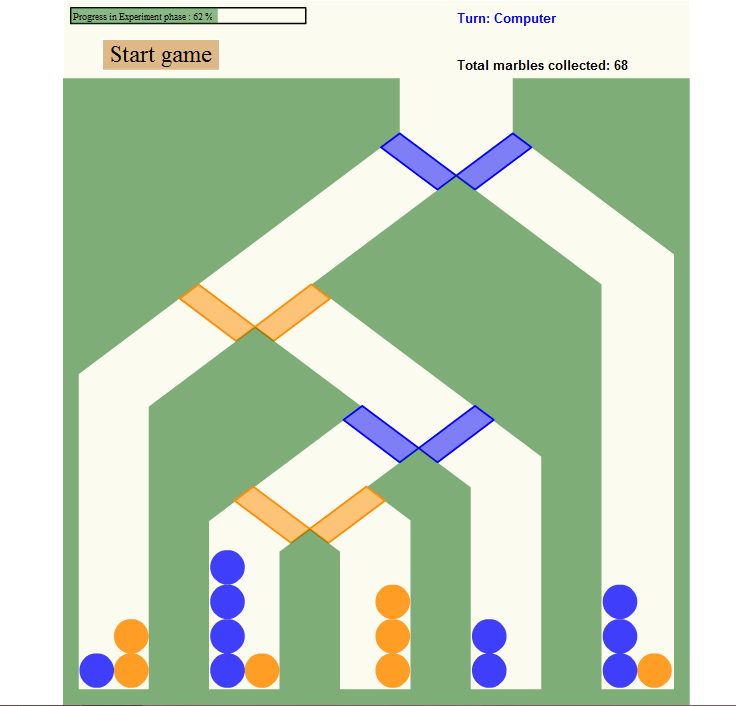}\\
& \\
\includegraphics[width=.22\textwidth]{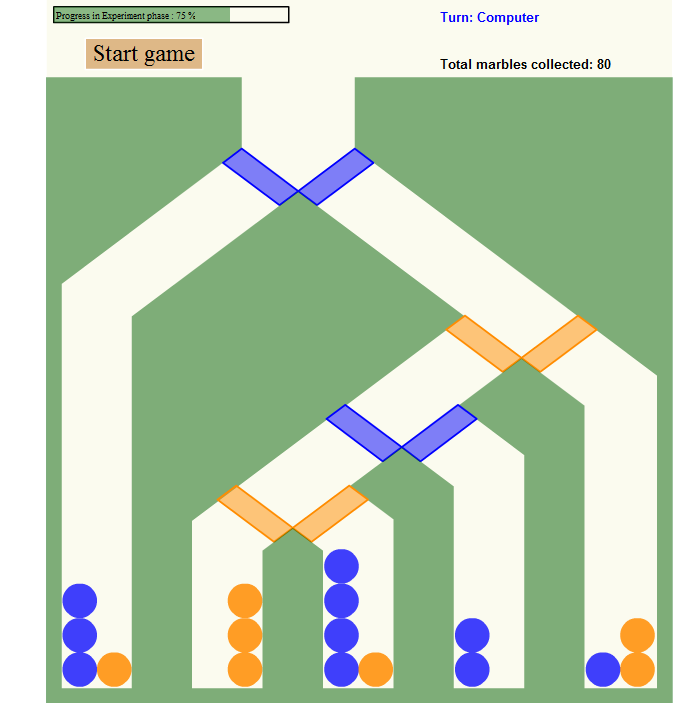} & \quad\includegraphics[width=.22\textwidth]{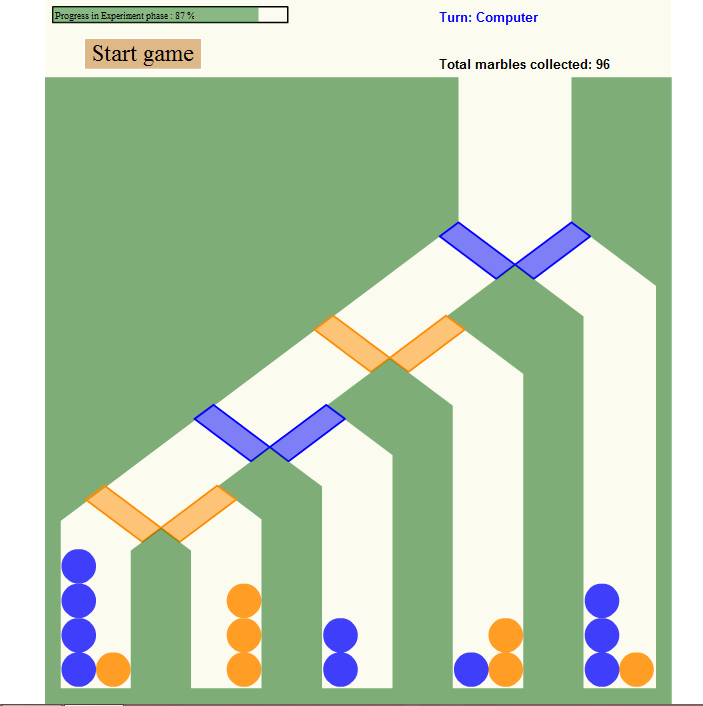} \up \up \up\\\\
\end{tabular}
\end{center}
\caption[]{Experimental games, various representations of game 1 of Figure~\ref{fig:maingames}} 
\label{fig:experimentgames}
\end{figure}

In each of the 8 rounds of the experimental phase, participants were confronted with all 6 games described in Section \ref{sec:games}. Different graphical representations of the same game were used in different rounds. As an example, Figure~\ref{fig:experimentgames} shows six visually different variations of game 1.

\section*{Appendix D: Answers to the final question}\label{sec:answer}

As mentioned in Section \ref{sec:design}, at the end of the experiment, each participant was asked the following question: ``When you made your choices in these games, what did you think about the ways the computer would move when it was about to play next?'' The participant needed  to describe   the plan he or she thought was followed by the computer on its next move after the participant's initial choice, in his or her own words. 

We found that one student who had made choices in the game that were consistent with FI reasoning, also provided an answer that suggested FI reasoning:

\begin{itemize}
\item[-] ``I first thought it would try to maximize the outcomes, taking into account that I would do the same. But I noticed that it did not always do that. Sometimes it did and sometimes it didn't. So after the break, I tried to maximize my outcomes, assuming the computer did the same, but if I noticed that the computer was not assuming that I would maximize my outcomes, I took a risk and I won a lot more.''
\end{itemize} 

Here follows a selection of answers provided by the other participants, which shows that participants might have given more importance to risk aversion and/or expected gains, rather than considering the outside option which the computer has already foregone.

\begin{itemize}
\item[-] ``I thought the computer took the option with the highest expected value. So if on one side you had a 4 blue + 1 blue marble and on the other side 2 blue marbles he would take the option 4+1= 2.5."
\item[-] ``It was going to take the turn with the highest reward, considering the risk. For example, when the computer can take a reward of 2 marbles instantly or choose to let the ball roll to an orange gate which has the potential of rewarding 4 marbles, the computer would go for the orange gate. With a difference of 1 marble between choices the computer is most likely to take the easiest way."
\item[-] ``It would choose for the safe 2/3 marble option instead of the dangerous 0/1 or 4 marble option."
\item[-] ``I made my choices based on how many marbles I could miss if the computer would turn left or right. In most cases I made the safe choice."
\item[-] ``My thoughts were about which most profitable route the computer would take, by looking at how many marbles the computer would get in comparison to me. If they were even or less then I think the computer would play safely and take the best and safest option available at that point."
\item[-] ``Look at the potential payoffs for blue in relation to the potential payoff for orange and check for probabilities."
\end{itemize}

\end{document}